\documentclass[prc,aps,preprint,showpacs,nofootinbib]{revtex4}
\usepackage{amssymb}
\usepackage[dvips]{graphicx}
\usepackage[english]{babel}
\usepackage{indentfirst}
\usepackage{amsxtra}
\usepackage{amsmath}
\usepackage{supertabular}
\usepackage{multirow}
\usepackage[mathcal]{eucal}

\newcommand{\btau}{\mbox{\boldmath $\tau$}}
\newcommand{\half}{\frac{1}{2}}

\newcommand{\slyt}{SLy5$_{\rm\,T}$\,}

\usepackage[usenames]{color}
\usepackage{ulem}

\begin{document}
\noindent 
\title {Tensor and tensor--isospin terms in the effective Gogny interaction }

\author{Marta Anguiano$^1$, Marcella Grasso$^2$,
 G. Co'$\,^{3,4}$, V. De Donno$\,^{3,4}$ 
and  A. M. Lallena$^1$}
\affiliation{
\mbox {1) Departamento de F\'\i sica At\'omica, Molecular y
  Nuclear,} \\ 
\mbox {Universidad de Granada, E-18071 Granada, SPAIN } \\
\mbox {2) Institut de Physique Nucl\'eaire, IN2P3-CNRS,} \\
\mbox {Universit\'e Paris-Sud, F-91406 Orsay Cedex, France}
\mbox {3) Dipartimento di Matematica e Fisica ``Ennio De Giorgi'',} \\
\mbox {Universit\`a del Salento, Via Arnesano, I-73100 Lecce, ITALY} \\
\mbox {4) INFN, Sezione di Lecce, Via Arnesano, I-73100 Lecce, ITALY }
}

\date{\today}
\bigskip

\begin{abstract}
We discuss the need of including tensor terms in the effective Gogny
interaction used in mean--field calculations. We show in one
illustrative case that, with the usual tensor term that is employed in
the Skyrme interaction (and that allows us to separate the
like--nucleon and the neutron--proton tensor contributions), we can
describe the evolution of the $N=28$ neutron gap in calcium isotopes.
We propose to include a tensor and a tensor--isospin term in
finite--range interactions of Gogny type. The parameters of the two
tensor terms allow us to treat separately the like--nucleon and the
neutron--proton contributions. Two parameterizations of the tensor terms have been chosen to 
reproduce different neutron single--particle properties in the $^{48}$Ca nucleus and the energy 
of the first $0^-$ state in the $^{16}$O nucleus. By employing these two parameterizations we 
analyze the evolution of the $N=14$, $28$, and $90$ neutron energy gaps in 
oxygen, calcium and tin isotopes, respectively.  We show that the combination 
of the parameters governing the like--nucleon contribution is crucial to
correctly reproduce the experimental (where available) or shell--model
trends for the evolution of the three neutron gaps under study.
\end{abstract} 

\vskip 0.5cm  
\maketitle

\section{Introduction}
\label{sec:int}

The presence of an electric quadrupole moment in the ground state of
the deuteron \cite{kel39,nor40} can be explained by including static
tensor terms in the microscopic nucleon--nucleon force as first
suggested by Rarita {\it et al.}~\cite{rar37,rar41a,rar41b}. This
procedure is today commonly adopted by all the modern microscopic
nucleon--nucleon interactions \cite{mac87,sto93,sto94,pud97,wir95}.
In a description of the nucleon--nucleon interaction based on a
meson--exchange picture \cite{mac87}, the strongest of the tensor
components, the tensor--isospin term, is dominated by the exchange
of a single pion. Since the pion is the lightest meson, this means
that the interaction range of the tensor--isospin term is the longest
one inside the nucleon--nucleon interaction.

For many years, tensor terms have not been considered in effective
interactions used in mean--field theories such as Hartree--Fock (HF) and
random--phase approximation (RPA), commonly used to
describe medium and heavy nuclei.  An exception to this is represented
by the semirealistic M3Y--P interactions~\cite{nak03,nak08}. These
interactions are based on the effective M3Y-Paris
interaction~\cite{ana83} which has been constructed to describe
inelastic nucleon--nucleus processes. The new M3Y--P interactions
are obtained by including a density--dependent zero--range term and by
modifying some of the force parameters. The tensor
and tensor--isospin terms of the M3Y--P3 and of the M3Y--P5 interactions
are the same as those used in the original M3Y--Paris interaction.

In almost all the existing parameterizations of the most used
effective interaction in HF and RPA self--consistent calculations, the
Skyrme interaction, the tensor term is neglected, even though a
zero--range tensor term was proposed in the original formulation of the force 
\cite{Sky56,Sky58}.  In the last years, tensor terms have been
included either on top of existing Skyrme parametrizations like the
SIII~\cite{bei75} and the SLy5~\cite{cha97,cha98a,cha98b} forces (see
Refs.~\cite{sta77,col07}) or by inserting them in the global fit
procedure producing new parametrizations of Skyrme
interactions~\cite{bro06,les07,ben09}.  In the following we shall
indicate as \slyt the parametrization introduced in
Ref.~\cite{col07} where tensor terms have been added on top 
of the SLy5 interaction.

Finite--range effective Gogny--like interactions are less used in HF
and RPA calculations than the Skyrme ones. As far as the tensor terms
are concerned, we find in the literature only few works where they
have been introduced in this type of forces. A first effort in this
direction was done by Onishi and Negele~\cite{oni78} who added a
tensor term to an effective force of finite range which was taken of
Gaussian form.  After the introduction of the Gogny force in
1980~\cite{dec80}, some further attempts of including a finite--range
tensor term have been done. Otsuka {\it et al.}~\cite{ots05} proposed
the GT2 force obtained by adding to the standard central channels of
the original Gogny force~\cite{dec80} a finite--range tensor--isospin
term of Gaussian form, and by refitting all the parameters. An
alternative procedure was proposed in Ref.~\cite{co11} where the D1ST
and the D1MT forces were constructed by adding on top
of the D1S~\cite{ber91} and D1M~\cite{gor09} parameterizations a
finite--range tensor--isospin term chosen to reproduce the energy of
the first 0$^-$ state in $^{16}$O in self--consistent HF plus RPA
calculations.

The inclusion of tensor terms in effective interactions allows us to have 
the same operator structure of the microscopic nucleon--nucleon interactions, and, moreover, 
it is necessary to describe observables related to both single particle (s.p.)
\cite{col07,col08,mor10} and collective properties of medium and heavy 
nuclei~\cite{cao09,don09,nes10,co11,ang11,co12b}.

The s.p. proton (neutron) energy gaps at $Z$ ($N$) = 8, 20, and 28 have been
investigated in Refs.~\cite{mor10,wan11} by using both non--relativistic and
relativistic HF techniques.  These studies showed that nuclear systems corresponding to 
$Z$ or $N =$ 8  and 20 are particularly suitable to study the neutron--proton tensor component 
of the effective interaction.

On the other hand, the s.p. proton (neutron) energy gaps along
isotonic (isotopic) chains may be sensitive to the like--nucleon
component of the tensor interaction. Experimental observations
indicate that the $N=14$ neutron gap in oxygen isotopes increases when
going from $^{16}$O to $^{22}$O. A similar behavior is found in
calcium isotopes for the $N=28$ neutron s.p energy gap which increases
from $^{40}$Ca to $^{48}$Ca. The shell--model calculations of
Ref.~\cite{sar10} describe this behavior and predict an analogous
increase of the $N=90$ neutron gap from $^{132}$Sn to $^{140}$Sn. In
that work, the previous effect has been attributed to the three--body
terms of the interaction. Here we show that, in the framework of
mean--field HF theory, the evolution of these three neutron gaps
strongly depends on the presence of the like--nucleon component of the
tensor term.

For the Skyrme interaction, several works exist in the literature
where both the neutron--proton and the like--nucleon tensor
contributions have been analyzed. An extensive study of the effects
generated by these two contributions has been carried out in Ref.~\cite{les07}. In the Gogny 
case, a detailed analysis where the two components are studied separately is still missing. As
already mentioned, in some recent works a finite--range tensor term
has been introduced only in the isospin dependent channel with a
single parameter to be chosen \cite{ots05,co11}. As it will be
discussed in the next section, this implies that the neutron--proton
and the like--nucleon contributions are proportional and have the same
sign, that is, they are both attractive or repulsive.  Considering
that realistic and semirealistic nucleon--nucleon forces include both
types of tensor terms (pure tensor and tensor isospin), we propose
here to take into account both terms also for the effective Gogny
interaction. This implies the introduction of a second parameter which
allows us to separately tune the neutron--proton and the like--nucleon
tensor contributions of the effective interaction.

The work is organized as follows. In Sec.~\ref{sec:compSk} we present
the physics case of the $N=28$ neutron gap in calcium isotopes. We
describe the experimental energies by using the Gogny D1ST and the
Skyrme \slyt interactions, and we show the need of including both
tensor and tensor--isospin terms in the Gogny interaction. In
Sec.~\ref{sec:neutron} we discuss
the implementation of the two finite--range tensor terms in the Gogny interaction, and we 
propose two possibilities for the choice of the parameters. In
Sec.~\ref{sec:res} we apply these two parameterizations of the
  tensor terms in the Gogny interaction to describe the evolution of the neutron gaps in
oxygen, calcium and tin isotopes. We compare our results with
experimental data (where available) and with the results of the HF
calculations carried on with the Skyrme interaction.  Finally, we draw
in Sec.~\ref{sec:con} our conclusions and we discuss the perspectives
of future applications of the Gogny plus tensor interaction.

\section{Neutron $N=28$ energy gap in calcium isotopes}
\label{sec:compSk}
It has been experimentally established that the $N=28$ neutron energy gap,
that is, the difference between the s.p. energies of the 2$p_{3/2}$ and
1$f_{7/2}$ neutron levels, increases when going from $^{40}$Ca to
$^{48}$Ca. The experimental situation is summarized in Fig.~7 of
Ref.~\cite{sor08} and indicates a change from a value of about
2.2 MeV in $^{40}$Ca to 4.8 MeV in $^{48}$Ca.

\begin{figure}
\begin{center}
\parbox[c]{16cm}{\includegraphics[scale=0.7,angle=0.0]
{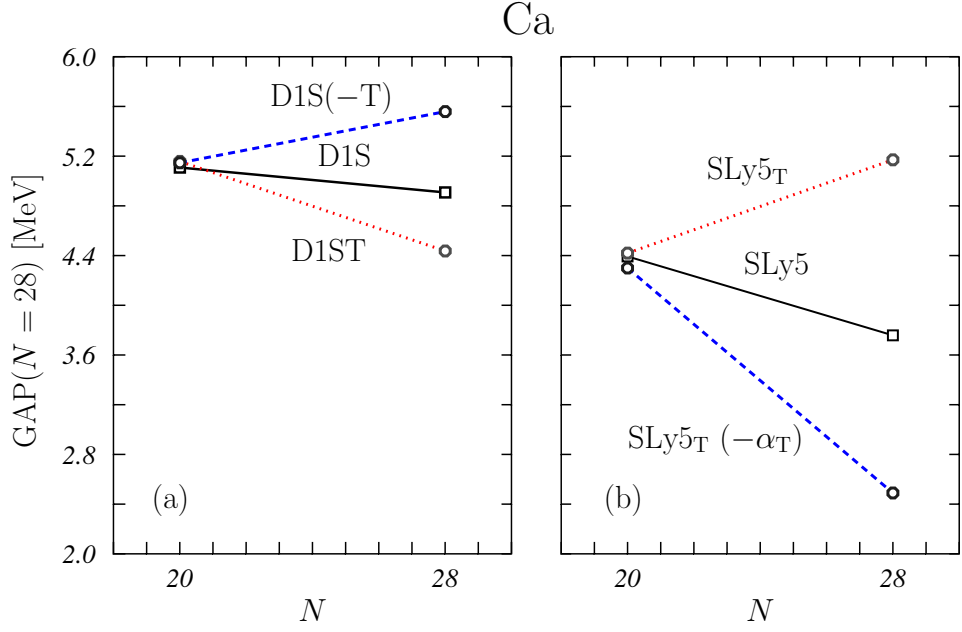}}
\caption{\small (Color online)(a) Neutron energy gap for $N=28$ in $^{40}$Ca and $^{48}$Ca nuclei
obtained with the Gogny interactions D1S (solid line) and D1ST (dotted line). 
The result obtained by changing the sign
of the tensor term in the D1ST case is shown by the dashed line. 
(b) Neutron energy gap for $N=28$ obtained 
with the Skyrme interactions SLy5 (solid line) and \slyt (dotted line). 
The result obtained by changing the sign of the parameter 
$\alpha_{\rm T}$ in the \slyt case is shown by the dashed line. 
}
\label{fig:skd1}
\end{center}
\end{figure}

We have calculated the evolution of this energy gap in the HF
framework by using Skyrme and Gogny interactions. Our results are
presented in Fig.~\ref{fig:skd1}.  In the panel (b) we show with
solid and dotted lines, respectively, the results obtained with the
Skyrme SLy5  and \slyt interactions. The values of the energy gaps we have obtained are, 
in general, larger than the experimental ones. Despite this deficiency, we observe that the
interaction without tensor terms, the SLy5, does not describe the
trend of the energy gap, which is slightly decreasing in this calculation. 
On the other hand, the result obtained with the \slyt
force, which includes tensor terms, shows an increasing behavior of
the energy gap.

The behavior of the energy gap is controlled by the like--nucleon term of
the tensor force. In the Skyrme interaction the contribution of tensor
components to the energy density of the system can be written as
\cite{col07,mor10} 
\begin{equation}
\Delta E_{\rm T}(r) = \half \alpha_{\rm T} 
\left[J^2_p(r) + J^2_n(r)\right] + \beta_{\rm T} J_p(r)
J_n(r) 
\,\,\,,
\label{eq:skyten}
\end{equation}
where $J_p(r)$ and $J_n(r)$ are the proton and neutron spin--orbit densities.
The parameters $\alpha_{\rm T}$ and  $\beta_{\rm T}$ rule, respectively, the 
like--nucleon and the proton-neutron terms of the tensor interaction. In the 
\slyt force \cite{col07} these parameters assume the values  -170~MeV~fm$^{5}$ 
and 100~MeV~fm$^{5}$, respectively. 

It is easy to show that the effect of the tensor interaction is
almost zero in spin-saturated nuclear systems, since the effect on the
$j=l+1/2$ s.p. level is canceled by that on the $j=l-1/2$ one. The
global effect would be exactly zero if the radial wave functions of
the two levels were the same.  Since calcium isotopes are
spin-saturated in protons, the like--nucleon tensor term
does not act on protons, and the neutron--proton
contribution is not active in the evolution of the neutron gap. The consequences of this 
in the excitation of magnetic states in calcium isotopes have been widely
discussed in \cite{co12b}.

The sensitivity of our results to the like--nucleon tensor term is
shown in the panel (b) of Fig.~\ref{fig:skd1} by the dashed line,
obtained by changing the sign of the parameter $\alpha_{\rm T}$.  This
modification leads to a decreasing energy gap going from $^{40}$Ca to
$^{48}$Ca. 

In the panel (a) of Fig.~\ref{fig:skd1} we show the results of HF calculations carried out 
with the Gogny interaction. The black full line indicates the result obtained with the D1S
force \cite{dec80} which does not contain tensor terms. The behavior
of the energy gap is analogous to that obtained with the Skyrme
interaction without tensor term. The dotted line shows the result
obtained with the D1ST interaction. In this case the behavior of the
energy gap is opposite with respect to the experimental one, and also
with respect to that obtained with the \slyt interaction. If the sign
of the parameter that determines the strength of the tensor term in
the D1ST interaction is changed, the results indicated by the blue dashed line in panel (a) 
of Fig.~\ref{fig:skd1} are obtained.  We remark that this operation on the D1ST force acts only
on the tensor--isospin dependent term and, therefore, changes both the
like--nucleon and unlike components of the tensor force. In this way,
nuclear properties depending on the neutron--proton tensor interaction
that are well described by the D1ST force are not any more
reproduced. For example, the energy of the first $0^-$ state in
$^{16}$O whose experimental value of 10.94 MeV was used to tune the
tensor force term in the D1ST interaction, appears at 14.48 MeV when
the sign of the total strength is changed. Evidently, a unique
tensor-isospin term in the D1S force is not able to reproduce
simultaneously both nuclear properties.

\section{Tensor terms and the Gogny interaction}
\label{sec:neutron}

The D1ST and D1MT interactions have been constructed
by adding a tensor-isospin term to the Gogny D1S and D1M interactions,
respectively~\cite{co11}.  The radial part of this term was based on
the analogous one in the microscopic Argonne V18
interaction~\cite{wir95}. Specifically, we have considered
\begin{equation}
v_{\rm Tt}(r) \,= \, v_{{\rm Tt}, {\rm AV18}} (r) \, \left[ 1\, -\, 
\exp \left(-\,b \, r^2 \right) \right] \, ,
\label{eq:vtensor}
\end{equation}
where the radial part of the Argonne V18 tensor
isospin term ~\cite{wir95}, $v_{\rm Tt, AV18}(r)$, 
has been multiplied by a function that simulates the
effect of the short-range correlations \cite{ari07}. Here 
$b$ is a free parameter.
The inclusion of this tensor-isospin term was done without changing
the values of the other force parameters but the strength of the
spin-orbit term. The values of the two free parameters, one for the
tensor and the other one for the spin-orbit term, have been chosen to
reproduce, in an iterative HF plus RPA calculation chain, the energy
of the first $0^-$ state and the s.p. energy gap between the
$1p_{3/2}$ and $1p_{1/2}$ neutron states, in $^{16}$O \cite{ang11}.

In the present work we use an expression for the tensor interaction
similar to that proposed by Onishi and Negele~\cite{oni78}
\begin{eqnarray}
\nonumber 
V_{\rm tensor} (r_1,r_2) &=& \left( V_{{\rm T}1} + V_{{\rm T}2} \,
P^{\tau}_{12} \right)\, S_{12} \,
\exp \left[ -(r_1-r_2)^2/\mu_{\rm T}^2 \right] \\
&=& \left[ \left(V_{{\rm T}1} + \half V_{{\rm T}2} \right) \,
+ \half V_{{\rm T}2} \, \btau(1) \cdot \btau(2) \right]\, S_{12} \,
\exp \left[ -(r_1-r_2)^2/\mu_{\rm T}^2 \right]
\label{eq:1fit}
\end{eqnarray}
where we have indicated with $P^{\tau}$ the usual isospin exchange
where $S_{12}$ and $\btau$ represent the usual tensor and
isospin Pauli operators. In the second line we have 
separated the pure tensor and tensor-isospin terms. In this approach
the radial part of the two independent tensor terms is identical, and
it has been chosen of Gaussian form. In our calculations we used
$\mu_{\rm T}=1.2$ fm, corresponding to the longest range of the D1S
interaction.

In this approach, the strength of the full tensor force is ruled by the two parameters 
$V_{{\rm T}1}$ and $V_{{\rm T}2}$. A calculation of the isospin matrix elements for the 
interaction (\ref{eq:1fit}) indicates that the strength of the force acting in
like--nucleon pairs is given by $V_{{\rm T}1}+V_{{\rm T}2}$,
while that between proton-neutron pairs is $V_{{\rm T}2}$. These
combinations of the parameters are, respectively, analogous to the
$\alpha_{\rm T}$ and $\beta_{\rm T}$ parameters of the Skyrme
interaction given in Eq. (\ref{eq:skyten}).

The two tensor terms in Eq. (\ref{eq:1fit}) have been added to the D1S
force without changing any other parameter value, including the
strength of the spin-orbit. In this way we are able to analyze
exclusively the effect of the tensor force.

In order to choose the values of the two free
parameters, $V_{{\rm T}1}$ and $V_{{\rm T}2}$, we have used two
observables. The first one is the energy difference between the
$1f_{5/2}$ and $1f_{7/2}$ s.p. neutron states in $^{48}$Ca. As already
discussed in Sect. \ref{sec:compSk}, this observable depends only on
the like--nucleon tensor contribution and therefore is ruled by
$V_{{\rm T}1}+V_{{\rm T}2}$.  We show in Table~\ref{tab:tuno} the
energy difference between these two s.p. states obtained for
various values of $V_{{\rm T}1}+V_{{\rm T}2}$. We have verified that
by changing $V_{{\rm T}1}$ and $V_{{\rm T}2}$ the result is the same
if the sum does not change. The experimental value of the energy
difference is 8.8 MeV~\cite{cot08}, therefore we have chosen $V_{{\rm
    T}1}+V_{{\rm T}2}=-20$~MeV.

\begin{figure}
\begin{center}
\parbox[c]{16cm}{\includegraphics[scale=0.8,angle=0.0]{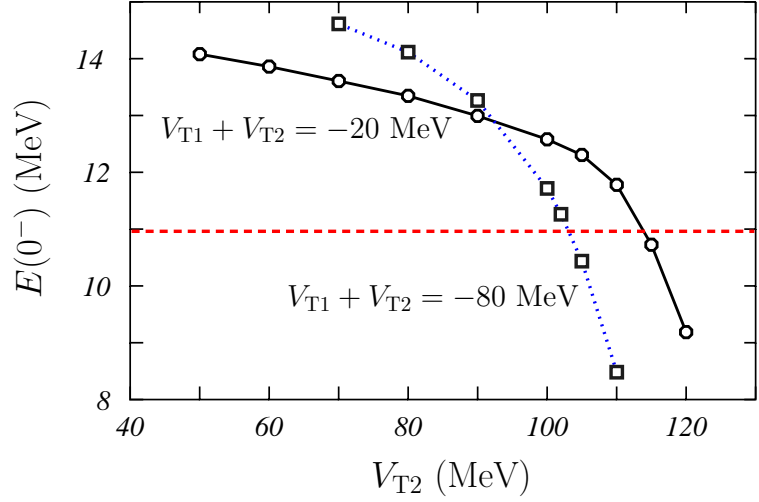}}
\caption{\small (Color online) 
Energy of the first 0$^-$ excited in $^{16}$O calculated in RPA as a
function of the $V_{{\rm T}2}$ parameter, Eq. (\ref{eq:1fit}), and keeping 
fixed $V_{{\rm T}1}+V_{{\rm T}2}$ at $-20$ MeV (solid line) and $-80$ MeV (dotted line). The 
horizontal dashed line indicates the experimental value.
}
\label{fig:e0}
\end{center}
\end{figure}

The second observable we have considered is the energy of the first $0^-$ state in
the $^{16}$O nucleus. In Ref.~\cite{co11} a large sensitivity of the energies of the 
$0^-$ states in doubly magic nuclei to the tensor-isospin term of the interaction was
observed.  We show in Fig.~\ref{fig:e0} the excitation energy of this
state calculated in the HF plus RPA approach for different values of
$V_{{\rm T}2}$. All the calculations shown by the black solid line have been carried out by 
using $V_{{\rm T}1}+V_{{\rm T}2}=-20$~MeV. For $V_{{\rm T}2}=115$~MeV we obtain for the energy
of the $0^-$ the value of 10.72 MeV, close to the experimental value
of 10.96 MeV \cite{led78}. This choice of $V_{{\rm T}2}$, together
with $V_{{\rm T}1}+V_{{\rm T}2}=-20$~MeV, implies $V_{{\rm T}1}=-135$
MeV. We label this parameterization D1ST2a. 

\begin{figure}
\begin{center}
\parbox[c]{16cm}{\includegraphics[scale=0.8,angle=0.0]{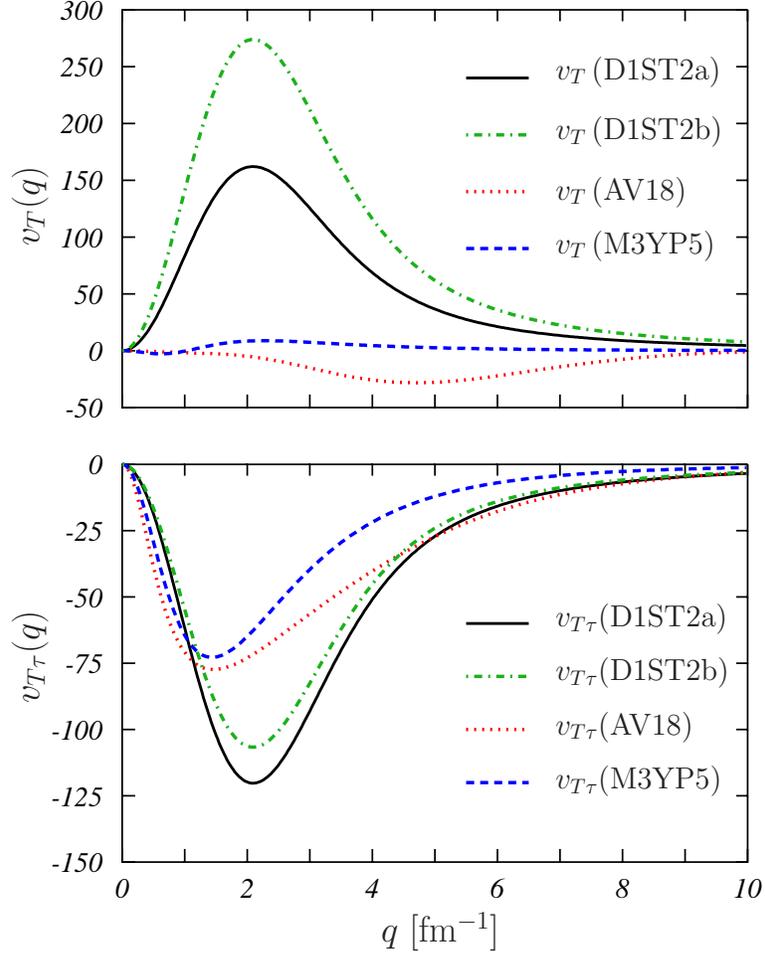}}
\caption{\small (Color online) 
Tensor (panel (a)) and tensor--isospin (panel (b)) terms of the D1ST2a and
D1ST2b parametrizations used in this work, compared with the analogous terms 
in the effective M3YP5 interaction and in the realistic one AV18.}
\label{fig:compare}
\end{center}
\end{figure}

In order to identify the general features of our results we have
implemented another parameterization of the tensor terms, which we call D1ST2b. In this case, 
we selected the like-nucleon part of the tensor force to reproduce the $N=28$ neutron gap 
increase from $^{40}$Ca to $^{48}$Ca as obtained in the HF calculation with the SLy5$_{\rm T}$ 
force. We obtained this results with the value of $V_{{\rm T}1}+V_{{\rm T}2}=-80$ MeV. As in the 
previous case, the other observable we have chosen to select the value of $V_{{\rm T}2}$ is
the excitation energy of the first $0^-$ state in $^{16}$O. The blue
dotted line in Fig.~\ref{fig:e0} indicates the value $V_{{\rm T}2}$ = $102$ MeV. 
In Fig.~\ref{fig:compare} we compare the two terms of the D1ST2a and D1ST2b tensor force with 
the analogous ones of the effective M3YP5~\cite{nak08} and 
microscopic AV18~\cite{wir95} interactions. The M3YP5 tensor isospin term is of the same order 
than that of our interactions. In the case of the $v_T$ term, M3YP5 presents an attractive part for small 
$q$, that becomes repulsive for $q>1$ fm$^{-1}$. It is interesting to notice that all the 
effective interactions have a repulsive $v_T$ term and an attractive $v_{T \tau} $ term. On the contrary, in the 
microscopic AV18 interaction both terms are attractive. This is an indication of the important role 
played by both short and long range correlations in modifying the interaction.

\section{Neutron gaps}
\label{sec:res}

The results we discuss in this section have been obtained in the HF
framework. Pairing correlations are not included in our calculations
since the nuclei we have considered, $^{16}$O, $^{22}$O, $^{40}$Ca,
$^{48}$Ca, $^{132}$Sn and $^{140}$Sn, have a well defined closed-shell
character.  The study has been conducted by comparing results obtained
by using interactions with (D1ST2a, D1ST2b and \slyt) and
without (D1S and Sly5) tensor terms.

\subsection{N=28 and N=90}
\label{su:gaps}

\begin{figure}
\begin{center}
\parbox[c]{16cm}{\includegraphics[scale=0.7,angle=0.0]{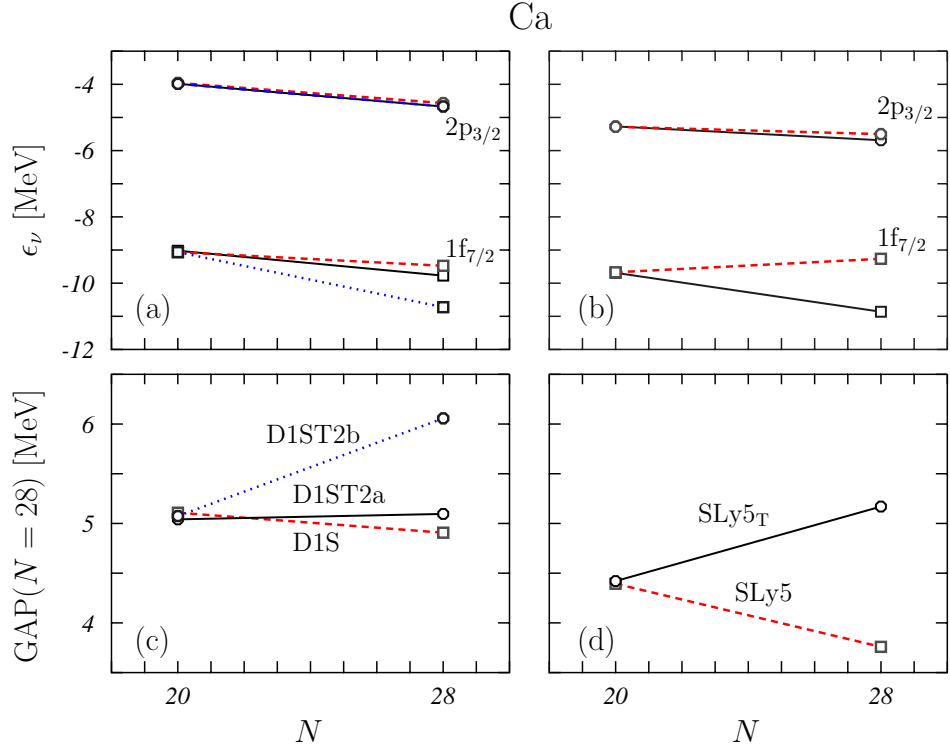}}
\caption{\small (Color online) 
Panel (a): Energies of the neutron s.p. levels around $N=28$ for the
nuclei $^{40}$Ca and $^{48}$Ca obtained with the 
D1S, dashed lines, D1ST2a, solid lines, and D1ST2b, dotted lines, forces.
Panel (c): evolution of the energy gap for $N=28$ obtained with the 
D1S, dashed line, D1ST2a, solid lines, and D1ST2b, dotted lines, forces.
Panel (b): the same as in (a) but for the  
SLy5, dashed lines, and \slyt, solid lines. Panel (d): the same as (c)
for the two Skyrme interactions.  
}
\label{fig:gap28}
\end{center}
\end{figure}

The gap evolution in calcium and tin isotopes, are rather similar. The
case $N=28$ involves the 2$p_{3/2}$ and 1$f_{7/2}$ neutron s.p. levels
in $^{40}$Ca and $^{48}$Ca.  The s.p. energies of these states are
shown in Fig.~\ref{fig:gap28}, for the Gogny (panel (a)) and Skyrme
(panel (b)) interactions. The corresponding gap values are shown in
panels (c) and (d) of the same figure.

The effects of the tensor on the energies of the 2$p_{3/2}$ state are
rather small, while those on the 1$f_{7/2}$ state are more evident,
producing a lowering of the energy value in $^{48}$Ca, much pronounced
in the case of the Skyrme and D1ST2b interactions. In both type of calculations 
(Skyrme and Gogny) only the presence of the tensor terms produces an increase of the gap,
in agreement with the experimental evidence \cite{sor08}. In the
shell--model calculations of Ref.~\cite{sar10} the energy of the 1$f_{7/2}$ level is lowered 
and that of the 2$p_{3/2}$  level is increased. This last effect is not present in our 
calculations.

\begin{figure}
\begin{center}
\parbox[c]{16cm}{\includegraphics[scale=0.7,angle=0.0]{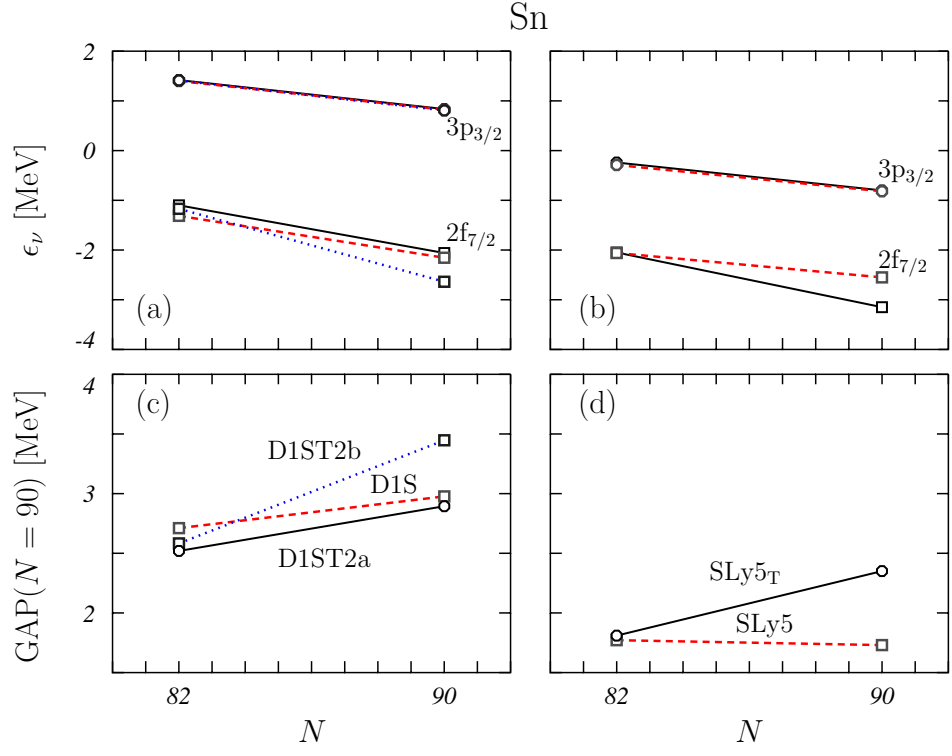}}
\caption{\small (Color online) The same as Fig. \ref{fig:gap28} for
  the case of the  $N=90$ neutron gap for the nuclei $^{132}$Sn and 
  $^{140}$Sn. The meaning of the lines is analogous to that of
  Fig. \ref{fig:gap28} with the obvious changes. 
}
\label{fig:gap90}
\end{center}
\end{figure}

The case $N=90$ involves similar s.p. states which differ from the
$N=28$ case only for the principal quantum numbers. The results
obtained are presented in Fig.~\ref{fig:gap90} and show behaviours
similar to those shown in Fig.~\ref{fig:gap28}. In this case, we found
an increase of the energy gap already in the D1S calculation. This
effect is enhanced by the inclusion of the tensor term and is more evident for the D1ST2b force. 
No experimental data are available for the $N=90$ gap, however shell model calculations carried 
out with microscopic interactions indicate an increase of the $N=90$ gap \cite{sor08}.


\subsection{N=14}

\label{su:gap14}

We discuss another case, in a different region of the nuclear chart,
where the experimental values of the s.p energies are
known. We consider the energy gap $N=14$ between the 2$s_{1/2}$ and
the 1$d_{5/2}$ neutron states in oxygen isotopes.  The experimental
value of this gap in $^{16}$O is 0.87 MeV \cite{vau72}. From the
study of the excited states in $^{21-23}$O nuclei through their
$\gamma$ decay, Stanoiu {\it et al.} \cite{sta04} deduced a value of
the energy gap of 4.11 MeV in $^{22}$O. This value is relatively large
and, for this reason, $^{22}$O can be considered a doubly magic
nucleus. This is also supported by the observation that the value of
the excitation energy of the first $2^+$ state in $^{22}$O is almost
twice that observed in the neighboring even--even nuclei. 

\begin{figure}
\begin{center}
\parbox[c]{16cm}{\includegraphics[scale=0.7,angle=0.0]{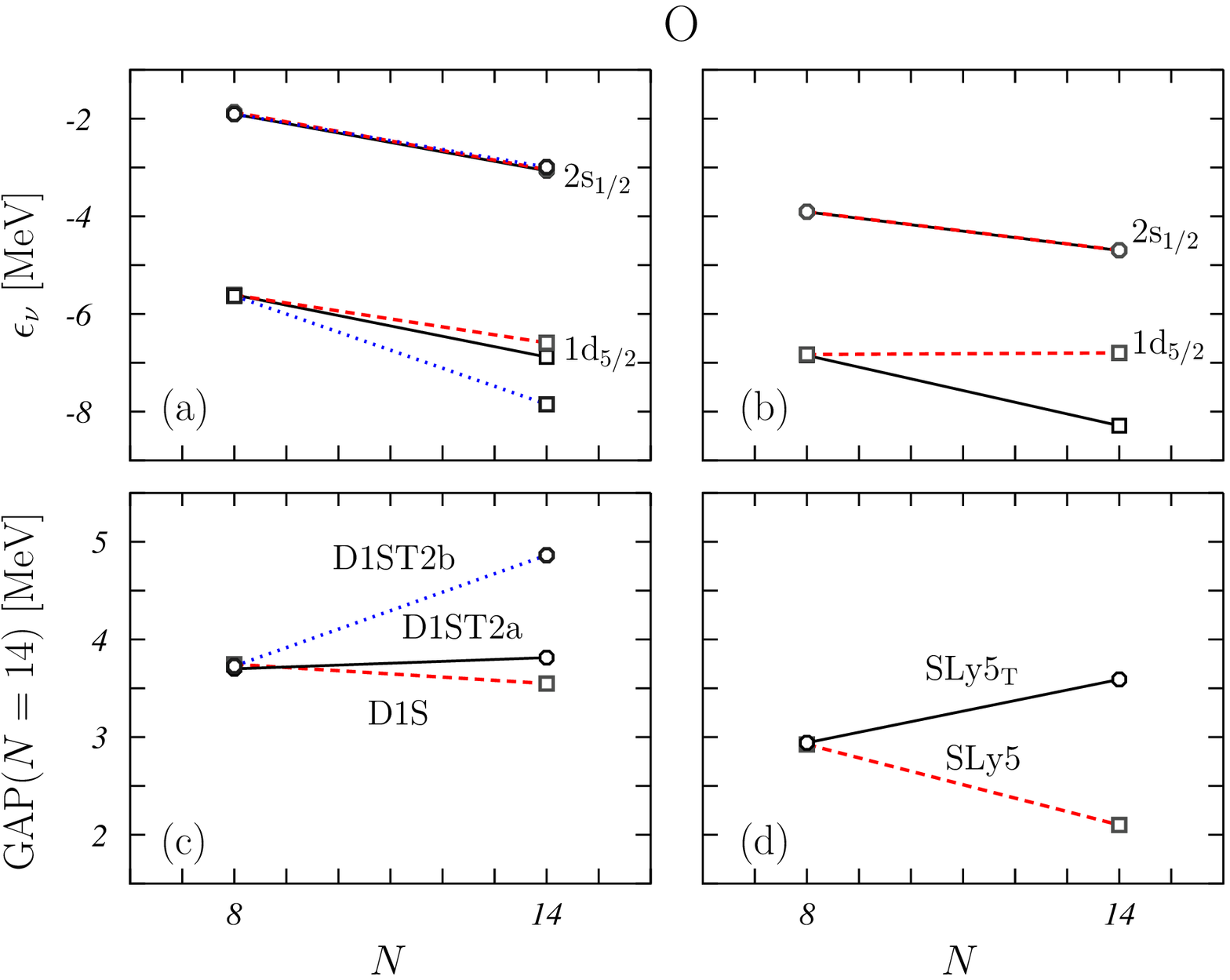}}
\caption{\small (Color online) 
  The same as Fig. \ref{fig:gap28} for
  the case of the  $N=14$ neutron gap for the nuclei $^{16}$O and 
  $^{22}$O. The meaning of the lines is analogous to that of
  Fig. \ref{fig:gap28} with the obvious changes.
}
\label{fig:gap14}
\end{center}
\end{figure}

The results of our calculations are shown in Fig.~\ref{fig:gap14}.
Also in this case the behavior found for the two types of
interactions, Gogny and Skyrme, are rather similar. The major effects
of the tensor terms of the force are present on the neutron 1$d_{5/2}$
s.p. energies which, in $^{22}$O are remarkably lower than those
obtained without tensor, mainly for the Skyrme
interaction and in the case of the D1ST2b force. This effect produces an
increase of the energy gap, even though the energies of the 2$s_{1/2}$
states remain unchanged. The results obtained with the D1S and SLy5
force show a decreasing gap.

\section{Conclusions}

\label{sec:con}

In this work, we have first pointed out the need of including in the
effective Gogny interaction two independent tensor terms acting
separately on like--nucleon and proton-neutron pairs. We have included
these two independent terms under the form of tensor and
tensor--isospin components to be added on top of effective Gogny
forces. To the best of our knowledge, only tensor--isospin terms have
been considered up to now for these finite--range interactions.
We have proposed two different parameterizations of the tensor force. In a first one, 
the strength of the like-nucleon part of the tensor force has been chosen to reproduce the 
experimental value of the splitting between the $1f_{7/2}$ and $1f_{5/2}$ neutron s.p. energies 
in $^{48}$Ca. In the second parameterization, the strength of this part of the interaction has 
been chosen to reproduce the neutron gap increase in $N=28$ going from $^{40}$Ca to $^{48}$Ca as 
obtained with the SLy5$_{\rm T}$ interaction. In both parameterizations the remaining term ruling 
the proton-neutron term has been selected to the energy of the first $0^-$ excited state in 
$^{16}$O

Using these parameterizations, we have calculated the neutron energy gap for $N=$
14, 28, and 90, in oxygen, calcium and tin isotopes respectively. Our
results show that both parameterizations reproduce the trend for the neutron gaps obtained with 
the Skyrme \slyt interaction, better in the case of the D1ST2b fit. This trend is
in agreement with the experimental behavior in oxygen and calcium
isotopes, and with the results of shell--model calculations in tin isotopes.

The inclusion of two tensor terms allows us to reproduce the
experimental trends of the neutron energy gaps in the isotope chains we have
investigated. This is our main result. From the quantitative point of
view it is evident that the two observables related to the
like-nucleon term of the tensor interaction are not compatible in HF
calculations. The parameterization D1ST2a built to reproduce the
s.p. splitting of the $f$ states in $^{48}$Ca produces the correct
behaviour of the neutron energy gap, but its value is quantitative too
small. Probably, a good quantitative description of these two
quantities requires to go beyond mean-field calculations, and to
consider explicitly the effects of the coupling between s.p. and
collective degrees of freedom.

We consider the present work as a step forward in the direction of
constructing a new parameterization of the Gogny interaction which
include tensor terms. We have the perspective of validating these new
tensor terms by using them in the description of observables where the
particle--like contribution of the tensor force is expected to play a
role, for example, in the excitation of unnatural parity states in
nuclei with neutron excess and with closed proton shells \cite{ang11}.
Of course, a more accurate fit would require to simultaneously modify
all the parameters of the Gogny force, especially the spin--orbit strength
which has a strong interplay with the tensor force, in both the
s.p. energies and the excitation of magnetic states.

\acknowledgments 
This work has been partially supported by the PRIN (Italy) {\sl
Struttura e dinamica dei nuclei fuori dalla valle di stabilit\`a}, by
the Spanish Ministerio de Ciencia e Innovaci\'on (Contract
Nos. FPA2009-14091-C02-02 and ACI2009-1007) and by the Junta de
Andaluc\'{\i}a (FQM0220).


\clearpage
\newpage
%
\begin{table}[htb]
\begin{center}
\begin{tabular}{cc}
\hline \hline
$V_{{\rm T}1}+V_{{\rm T}2}$ \,\,\,   & $\epsilon_{\nu}$(1$f_{5/2}$) -$\epsilon_{\nu}$(1$f_{7/2}$) \\
\hline 
  0.0    &   8.3  \\
 -5.0    &   8.5  \\
-10.0    &   8.6   \\
-20.0    &   8.9    \\
-30.0    &   9.3    \\
-40.0    &   9.7    \\
-50.0    &  10.0      \\
-60.0    &  10.4 	    \\
-70.0    &  10.8	     \\
-80.0    &  11.1   \\
\hline \hline
\end{tabular}
\end{center}
\caption{Difference between the energies of the $1f_{5/2}$ and
  $1f_{7/2}$ s.p. neutron states in $^{48}$Ca for some values of 
$V_{{\rm T}1}+V_{{\rm T}2}$. The experimental
value is 8.8 MeV~\cite{cot08}. All the quantities are expressed 
in MeV.} 
\label{tab:tuno}
\end{table}

\clearpage
\newpage
%
%
%
%
%
%
%

%

\end{document}